\documentclass[12pt]{article}
\usepackage{graphicx}

\author{I. Filikhin, V. M. Suslov
and  B. Vlahovic\thanks{This work was supported
by NSF CREST award HRD-0833184 and NASA award NNX09AV07A.
The numerical calculations were performed by High Performance Computing Center of the North Carolina State University}\\
Department of Physics, \\
North Carolina Central University, Durham, NC 27707, USA
}

\title{Spin-flip doublets of $^9$Be  spectrum within a cluster model
}

\begin{document}
\maketitle


\begin{abstract}
The structure of the $^9$Be low-lying spectrum is studied within the cluster model
$\alpha+\alpha+n$. In the model the total orbital momentum is fixed for each
energy level. Thus each level is determined as a member of the spin-flip
doublet corresponding to the total orbital momentum ($L^\pi=0^+, 2^+,4^+, 1^-, 2^-,3^-, 4^-$) of the
system.  The
Ali-Bodmer potential (model E) is applied for the $\alpha\alpha$ interaction.
We employ a local $\alpha n$ potential which was constructed to reproduce the $\alpha-n$
scattering data.
The Pauli blocking is
simulated by the repulsive core of the $s$-wave components of these potentials.
Configuration space Faddeev equations
 are used to calculate the energy of the bound state ($E_{cal.}$=-1.493 MeV v.s.
$E_{exp.}$=-1.5735 MeV) and resonances.
A variant of the method of analytical continuation in the coupling constant is applied
to calculate the energies of low-lying levels.
Available $^9$Be spectral data are satisfactorily reproduced by the proposed model.
\end{abstract}


\section{\label{Itr} Introduction}

The  $^9$Be nucleus can be modeled as a typical cluster nuclear system with the neutron-halo structure.
The three cluster model $\alpha+\alpha+n$ allows us to describe the $^9$Be low-lying spectrum qualitatively \cite{Tang}.
In the last years the cluster calculations for the $^9$Be spectrum have been performed using different techniques and with
different inter-cluster potentials \cite{Arai,T2006,T2007,ARJGF}.
The main problem of the cluster models is the adequate description of inter-cluster interactions. In particular, for an $\alpha+\alpha+n$ system, it
is important to take into account the Pauli blocking for both the $\alpha\alpha$ and $\alpha n$
interactions. There exist two different methods of approximation for considering this problem.  The first uses the process of elimination on the Pauli forbidden states caused by the potentials used. This procedure realizes the non-local potentials \cite{T2007, Hiyama}.
The second uses the potentials which have a repulsive core to simulate the Pauli blocking \cite{SBB,Zhukov1993}.
In this work we apply the second method of approximation.
The inter-cluster potentials can be phenomenologically constructed  by experimental data for the $\alpha\alpha$ and $\alpha n$ scattering.
We used the known Ali-Bodmer potential (model E) \cite{ABo66} for the $\alpha\alpha$ interaction.
The $\alpha n$ potential was constructed to reproduce the results of R-matrix analysis for the  $\alpha n$ data  \cite{Noll2007}. In this way we adjusted the
parameters of the $\alpha n$ potential taken from \cite{Zhukov1993,FJR}.
Our goal is to obtain a new theoretical description for the  $^9$Be spectrum. Note  that  reliable $^9$Be experimental data is restricted by the first
several levels. In our model the total orbital momentum is fixed for each
level and, taking into account the spin of the neutron, each of the levels are determined as members of the spin-flip
doublet corresponding to the total orbital momentum of the
system. We consider the states of the  $\alpha+\alpha+n$ system with $L^\pi=0^+, 2^+,4^+, 1^-, 2^-,3^-, 4^-$.
To calculate the energies of the low-lying levels, we applied
the configuration space Faddeev equations \cite{Faddeev}. According to the proposed model, the partial decomposition of the Faddeev components of the total wave function was performed in the $LS$ coupling scheme.

We found that the $^9$Be spectral data are well reproduced within the proposed model. Thus we present a new classification of the $^9$Be low-lying spectrum as a set of spin-flip doublets.  A comparison of our results
with the results obtained by different theoretical approaches supports our approximation as well.
Predictions for the $^9$Be spectrum are also presented.

\section{\label{Form} Formalism}

The Faddeev equations in coordinate space
\cite{Faddeev} are used for description of the $^9$Be nucleus, considered as a three-body  $\alpha\alpha n$ system.
The general form of the equations is as follows:
\begin{equation}
\{H_0+V^s_{\gamma}(\vert\vec{x}_{\gamma}\vert)+
\sum_{\beta =1}^3V^{Coul.}_{\beta}(\vert\vec{x}_{\beta}\vert)-E\}
\Psi_{\gamma}(\vec{x}_{\gamma},\vec{y}_{\gamma})
=-V^s_{\gamma}(\vert\vec{x}_{\gamma}\vert)
\sum_{\beta\ne\gamma}\Psi_{\beta}(\vec{x}_{\beta},\vec{y}_{\beta}),
\label{F0}
\end{equation}
where $V^{Coul.}_{\beta}$ is the Coulomb potential between the particles
belonging to the pair $\beta$ and $V^s_\gamma$ is the short-range
pair potential in the channel $\gamma$, ($\gamma$=1,2,3).
$H_0=-\Delta_{\vec{x}_{\gamma}}-\Delta_{\vec{y}_{\gamma}}$ is the
kinetic energy operator, $E$ is the total energy, and $\Psi$ is the wave
function of the three-body system. $\Psi$ is given as a sum over
three Faddeev components, $\Psi =\sum^3_{\gamma=1}\Psi_\gamma$.
For the $\alpha\alpha\Lambda$ system including two identical particles, the coupled set of the
Faddeev equations is written as:
\begin{equation}
\begin{array}{l}
(H_0+V_{\alpha n}+V^{Coul.}-E)W=-V_{\alpha n}(U+P_{12}W),\\
(H_0+V_{\alpha \alpha}+V^{Coul.}-E)U=-V_{\alpha \alpha}(W+P_{12}W),\\
\end{array}
\label{F}
\end{equation}
\noindent where $U$ is the Faddeev
 component corresponding to the rearrangement channel
$(\alpha\alpha)-n$ and $W$ corresponds to the rearrangement
channel $(\alpha n)-\alpha$. The total wave function is
expressed by the components $U$ and $W$ with the relation $\Psi = U + (1
+ P_{12})W$. $P_{12}$ is the permutation operator for the
 $\alpha$ particles (particles 1,2),
$V_{\alpha\alpha}$ and $V_{\alpha n}$ are nuclear
potentials of $\alpha\alpha$ and $\alpha n$ interactions,
respectively. $V^{Coul.}$ is the potential of Coulomb interaction between
the $\alpha$ particles,  The total orbital angular momentum is given by $\vec
L={\vec\ell}_{\alpha\alpha}+{\vec\lambda}_{(\alpha\alpha)-n}=
{\vec\ell}_{\alpha n}+{\vec\lambda}_{(\alpha n)-\alpha}$,
where ${\ell}_{\alpha\alpha}$ (${\ell}_{\alpha n}$) is the
orbital angular momentum of $\alpha$$\alpha$ (pair of
${\alpha n}$) and ${\lambda}_{(\alpha\alpha)- n}$
(${\lambda}_{(\alpha n)-\alpha}$) is the orbital angular
momentum of a neutron ($\alpha$ particle) relative to
the center of mass of the $\alpha\alpha$ pair and $\alpha n$ pair, respectively.

Possible combinations of relative momenta
${\ell}_{\alpha\alpha}$,
  ${\lambda}_{(\alpha\alpha)-n}$
 and ${\ell}_{\alpha n}$,
 ${\lambda}_{(\alpha n)-\alpha}$ for the system with total orbital momentum $L$ and parity $\pi$
written as a block matrix are:
$$M(L^\pi)=\left(
  \begin{array}{ccccccc}
  ( & \{{\ell}_{\alpha n}\}& ) & (&\{ {\ell}_{\alpha\alpha}\} & ) \\
  ( & \{{\lambda}_{(\alpha n)-\alpha} \}& ) & ( &\{{\lambda}_{(\alpha\alpha)- n}\}& ),
  \end{array}
\right)
$$
where each block represents all quantum
numbers taken into account. For example, the combinations corresponding to the $0^+$ state
can be given in the form:
$$
M(0^+)=\left(
  \begin{array}{cccccccccccc}
( & 0 & 1 & 2 &\dots & ) & ( & 0 & 2 & 4&\dots & ) \\
( & 0 & 1 & 2&\dots & ) & ( & 0 & 2 & 4 &\dots& ).
\end{array}
\right)
$$
A more detailed description of this formalism is given in Ref. \cite{FGS04}, applied to  the
cluster system $\alpha \alpha \Lambda$.
For the system $\alpha \alpha n$, the spin-orbit coupling between the $\alpha$-particle and the neutron is not negligible and we
 implement the  spin-orbit part of the $\alpha n$ interaction into the formalism.
 The $\alpha n$ potential
 is written as  a sum of the central and the spin-orbit parts: $V_{\alpha n}=V^c_{\alpha n}+V^{so}_{\alpha n}$.
 For the $\alpha \alpha n$ system in the $LS$ basis
the matrix elements of the spin-orbital potential $V^{so}_{\alpha n}$ are given by the following form:
 \begin{equation}
 V^{so}_{\alpha n}(r)=
\frac{2L+1}2\sum_{j=l \pm 1/2}(2j+1)
\left \{ \begin{array}{ccc}
    J & L & 1/2  \\
    l & j & \lambda
    \end{array} \right \}^2
\end{equation}
\[
\times (j(j+1)-l(l+1)-3/4)v_{so}(r),\]
where $J$ is the total three-body angular momentum, $L$ is the total three-body orbital momentum,
$j$ and $l$ are total orbital momenta of the ${\alpha n}$ pair (the spin of the pair is equal to $\frac12$),
$\lambda$ isthe  orbital momentum of the $\alpha$-particle with respect to the center of the ${\alpha n}$ pair, and
$v_{so}(r)$ is a coordinate part of the $\alpha n$ spin-orbit potential.

The configurations of the angular momenta corresponding to the
$L^\pi$ states taken into account are represented below:

\noindent{\bf $2^+$ state}
$$
M(2^+)=\left(
  \begin{array}{cccccccccccccccc}
 ( & 0 & 1 & 1 & 2 & 2 & 2 & ) & ( & 0 & 2 & 2 &  2 &  4 & 4 &) \\
 ( & 2 & 1 & 3 & 0 & 2 & 4 & ) & ( & 2 & 0 & 2 & 4 & 2 & 4 &)
  \end{array}
\right),
$$
\noindent{\bf $4^+$ state}
$$
M(4^+)=\left(
         \begin{array}{ccccccccccccccccc}
( & 0 & 1 & 2 & 2 & 3 & 4 & ) & ( & 0 & 2 & 4 &  2 & 4 & 4 &) \\
( & 4 & 3 & 2 & 4 & 1 & 0 & ) & ( & 4 & 2 & 2 & 4  & 0 & 4 &)
         \end{array}
       \right),
$$
{\bf$2^-$ state}
$$
M(2^-)=\left(
         \begin{array}{ccccccccccccccc}
( & 1 & 2 & 2 & 3 & 3& 4 & ) & (  & 2 & 2 & 4 & 4& ) \\
( & 2 & 1 & 3 & 2 & 4 & 3 & ) & ( & 1 & 3 & 1 & 3& )
         \end{array}
       \right),
$$
{\bf $4^-$ state}
$$
M(4^-)=\left(
         \begin{array}{ccccccccccccccc}
( & 1 & 2 & 2 & 3 & 4 & 4 & ) & ( & 2 & 2 & 4 &  4 &) \\
( & 4 & 3 & 5 & 2 & 1 & 3 & ) & ( & 3 & 5 & 1 & 3 &)
         \end{array}
       \right),
$$
{\bf $3^-$ state}
$$
M(3^-)=\left(
         \begin{array}{cccccccccccccccc}
( & 0 & 3 & 2 & 1 & 1 & 4 & ) & ( & 0 & 2 & 2 &  4 &  4 &) \\
( & 3 & 0 & 1 & 2 & 4 & 1 & ) & ( & 3 & 1 & 3 & 1  & 3  &)
         \end{array}
       \right),
$$
{\bf $1^-$ state}
$$
M(1^-)=\left(
         \begin{array}{cccccccccccccccc}
( & 0 & 1 & 1 & 2 & 2 & 3 & ) & ( & 0 & 2 & 2 &  4 & 4 &) \\
( & 1 & 0 & 2 & 1 & 3 & 2 & ) & ( & 1 & 1 & 3 &  3 & 5 &)
         \end{array}
       \right).
$$

\section{\label{Pot} Potentials}

Nuclear $\alpha\alpha$
interaction is given by the phenomenological
Ali-Bodmer (AB) potential \cite{ABo66}. This potential has the following form:
 $V_{\alpha\alpha}(r) =  \sum_{l=0,2,4}
V^l_{\alpha\alpha}(r)P_l,$ where $P_l$ is a projector onto the state
of the $\alpha \alpha$ pair with the orbital momentum $l$. The
functions $V^l_{\alpha\alpha}(r)$ have the form of one or two range
Gaussians:
\begin{equation}
\label{pot} V^l_{\alpha\alpha}(r) =
V^l_{rep}\exp(-\beta^l_{rep}r)^2-V^l_{att}\exp(-\beta^l_{att}r)^2\;\;.
\end{equation}
The $s$-wave component
$V^0_{\alpha\alpha}(r)$ has a strong repulsive core which simulates
Pauli blocking for the $\alpha$'s at short distances.
There are different sets of the  parameters for partial components
$V^l_{\alpha\alpha}(r)$ \cite{ABo66}.
The parameters of the version "e" of the Ali-Bodmer potential (ABe) are
given in Table~\ref{tab:1}.

\begin{figure}
\includegraphics[width=120.0mm]{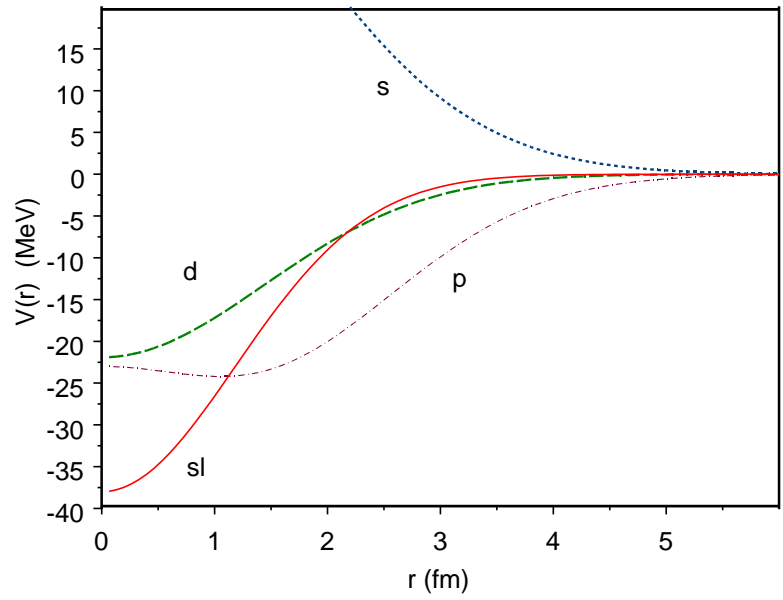}
\caption{\label{Filikhin_fig:1}
Partial ($s,p,d$) and spin-orbit ($sl$) components of the $\alpha-n$ potential.
}
\end{figure}

The $\alpha n$ interaction in $s$, $p$ and $d$ states is taken into account.
The $p$ and $d$-wave components include central and spin-orbit parts:
\begin{equation}
V^{jl}_{\alpha n}(r) = v^l_{c}(r)+({\vec s,\vec l})v_{so}(r),
\label{an}
\end{equation}
where $\vec j=\vec l+ \vec s$, $\vec s$ is the spin of the neutron.
The coordinate dependencies of the components havethe  form of one or two range Gaussians.
The $s$-wave component has a repulsive core \cite{SBB,Zhukov1993,FJR} to simulate the $s$ state Pauli exception for $\alpha-n$.
The $d$-wave component of the $\alpha n $ potential was taken from Ref. \cite{FJR}.
In this work we propose a modification for the $p$-wave and spin-orbit components of the potential given in \cite{FJR}.
The parameters of our potential are listed in Table~\ref{tab:1}.
The goal of the modification is to reach a better description for the $\alpha-n$ scattering data.
 For experimental data, we used the results of the R-matrix analysis of the data from \cite{Noll2007}. In Fig.~\ref{Filikhin_fig:1}
the phase shifts for the $\alpha-n$ scattering are given to compare
our obtained results with the proposed potential and the R-matrix fit.
We have obtained strong agreement between the two results for neutron energies up to 4-5~MeV.

\section{\label{Meth} Methods}
The bound state problem based on the configuration space Faddeev equations (\ref{F0}) for the
$\alpha\alpha n$ system (Eq. (\ref{F})) is solved numerically by applying the
finite difference approximation with spline
collocation method \cite{FGS04, BSSV96}. For calculation of the
eigenvalues, the method of inverse iterations is used.
To estimate the energies and widths of low-lying resonance states, we applied the method of analytical continuation in the coupling constant
\cite{KKH}.  A variant of this method with an additional
non-physical three-body potential is used. The strength parameter
of this potential is considered as a variational parameter for the
analytical continuation of the bound state energy into the complex plane
\cite{KK,FSV05,FSV09}. This potential, considered as a perturbation to the
corresponding three-body hamiltonian, is added to the left hand side
of the equations (\ref{F}). The three-body potential has the form:
\begin{equation}
V_3(\rho) = -\delta\exp(-b\rho^2),
\label{3bf}
\end{equation}
where $b$ is a range parameter of this potential ($b$=0.008 $fm^{-2}$)  and $\delta$ is a
strength parameter (variational parameter), $\delta \geq$0.
$\rho^{2}=x_{\alpha}^{2}+y_{\alpha}^{2}$, where
$x_{\alpha},y_{\alpha}$ are the mass scaled Jacobi coordinates
($\alpha=1,2$) \cite{FSV}. For each resonance there exists a
region $|\delta |\geq |\delta_0 |$ where a resonance becomes a bound
state. In this region we obtain the 2$N$ three-body bound state energies
corresponding to 2$N$ values of $\delta$.
2N is the number of points used in the real energy plane for construction of the Pad\'{e}
approximation.
The continuation of the
energy into the complex plane is carried out by means of the Pad\'{e}
approximation: $
   \sqrt{-E}=\frac{\sum_{i=1}^Np_i\xi^i}{1+\sum_{i=1}^Nq_i\xi^i}
$ where $\xi=\sqrt{\delta_0-\delta}$ and $p_i$ and $q_i$ are calculated parameters. The Pad\'{e} approximation for
$\delta=0$ gives the energy and width of the resonance:
$E(\delta=0)=E_r+i\Gamma/2$. The accuracy of the Pad\'{e}
approximation for resonance energy and width depends on a few parameters, the distance from the scattering threshold, the accuracy of calculation for
bound states and determination of $\delta_0$. Calculated resonance energy can depend on the order $N$ of the Pad\'{e}
aproximants used (for example, see  Ref. \cite{Car}).

\section{\label{Cal} Calculations}

The bound state of the $\alpha\alpha n$ system, having negative ("natural" \cite{T2007})) parity with
$J^\pi=(\frac32)^{-}$ and  $L^\pi$=1$^-$,  is associated with the ground
state of the $_{\Lambda}^{~9}$Be hypernucleus. The experimental value for this state is 1.5735~MeV \cite{tunl}.
This state is lower member of the spin-flip doublet with orbital momentum $L$=1.
The upper member of this doublet is the resonance state $J^\pi=(\frac12)^{-}$ with energy 1.21$\pm 0.12$~MeV.
Calculated values (~0.9 MeV) for this doublet are close to the experimental data. The calculated spin-doublet spacing
is about 2.4 MeV, whereas the experimental value is about 2.68$\pm 0.12$ MeV.
In Table. \ref{Filikhin_tab:2} we give the $^{9}$Be ground state binding energy calculated
for various orbital momentum configurations.
Orbital momentum configurations $l^\pi=0^+$ and $ 2^+$ of the core nucleus $^8$Be contribute significantly to the $^{9}$Be ground state energy.
A weakly bound state of the $\alpha \alpha n$ system  is possible if the $l^\pi=0^+$ configuration is taken into account. Addition of the
 $l^\pi=2^+$ configuration of the $\alpha \alpha$ pair gives the final value for the binding energy.  Meanwhile, the
 contribution of the configuration $l^\pi=4^+$ is relatively small.

Results of our calculation for the low-lying spectrum of  $^{9}$Be are given
in the third column of Table \ref{Filikhin_tab:3}. The next two columns include the results of the
Refs. \cite{Arai, ARJGF,P}. Experimental data for $^9$Be ($T=1/2$) is given in the last column of this table.
Each energy level is classified as a member of the spin-flip
doublet corresponding to the total orbital momentum $L^\pi=0^+, 2^+,4^+, 1^-, 2^-,3^-, 4^-$ of the $\alpha \alpha n$
system. Our model's predicted results correlate well with the experimental data \cite{tilley}. In our opinion, the neutron-halo structure of
the  $^{9}$Be is the reason for the  possibility of classification.

\begin{figure}
\includegraphics[width=130.0mm]{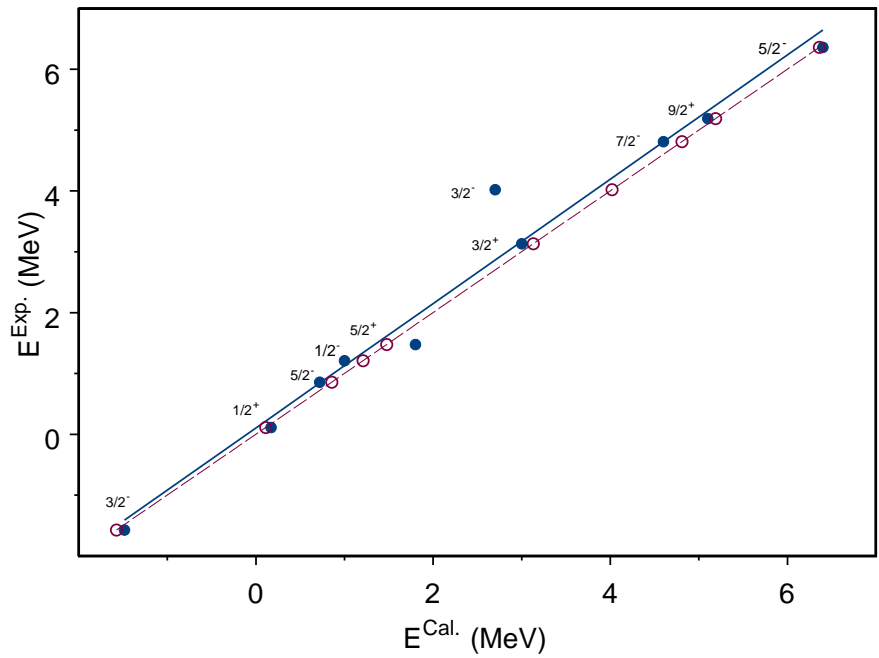}
\caption{\label{Filikhin_fig:5}
Correlation between calculated (Cal.) and experimental (Exp.)  spectrum of $^9$Be (solid dots).
Solid line is the root mean square fit for the correlation.
The dashed line shows the ideal situation when the calculated values coincide with experimental data (open dots).
Total momentum of each level is shown.
}
\end{figure}

To give overall comparison of our experimental results,  we present
a
relation  between the calculated  and experimental spectrum of $^9$Be in Fig. \ref{Filikhin_fig:5}.
In this figure the solid line indicates the root mean square fit for this correlation. The fitted line is close to
the line (dashed line in Fig.\ref{Filikhin_fig:5})  showing  the ideal situation when calculated values coincide with experimental data. These two lines are practically identical. The slight angle difference between the lines is due to
a relatively large disagreement between calculated and experimental values
for the  $J^\pi=3/2^-$ ($L^\pi=2^-$) and  $J^\pi=5/2^+$ ($L^\pi=2^+$) states.
Note that other cluster calculations \cite{Arai,ARJGF} (see Table \ref{Filikhin_tab:2}) demonstrate
the same disagreement with the experimental data. We note that
there is experimental evidence \cite{Prezado} for the wide $J^\pi=(\frac32)^{-}$ resonance  with energy of 3.4$\pm$0.5 MeV that is closer to the calculated values.

Our calculations strongly agree with previous $^9$B
calculations \cite{Arai,P}. In particular, we confirm that the $J^\pi=7/2^+$ and $J^\pi=9/2^-$ resonances
have the excited energies about 10~MeV and 8~MeV, respectively.
Energy of the resonance $J^\pi=7/2^-$ as an upper member of the level $L^\pi=4^-$ was predicted with approximately an excitation  energy of 12 MeV.

The mirror nucleus  $^9$B can also be considered using the present cluster model after the replacement of neutron by proton. Obviously, this replacement must include the changes in mass and potential.
To evaluate the effect of the potential replacement, we calculated the first energy levels of the $\alpha\alpha p $ system with the $\alpha$-$n$
potential which was used for the $\alpha\alpha n$ calculations.
Results of our calculations for several levels are given in Table \ref{Filikhin_tab:2}. We compare these results with experimental data for   $^9$Be and  $^9$B.
The Coulomb energy $\Delta_c$ is calculated as the energy difference between corresponding levels of $^9$Be and  $^9$B, measured from the $\alpha+\alpha+n$ and $\alpha+\alpha+p$ thresholds, respectively. This energy includes the result of the change in mass which has to be excluded from this difference.  Note that the mass difference of proton and neutron is about 1.293 meV.
From Table \ref{Filikhin_tab:3} it is clear that the $\alpha$-$n$
potential used in the $\alpha\alpha p $ system leads to systematical overbinding (about 0.25 MeV and more) for the $\alpha\alpha p $ system relative to the experimental data.
The Coulomb energy $\Delta_c$ is slightly less (about 0.15 MeV) than the experimental value which is in conflict with this overbinding. One can conclude that the $\alpha p$ potentials has to differ from the $\alpha$$n$ potential in both strength and range parameters.
Nevertheless, the obtained results qualitatively reflect a relation between the $^9$B low-lying levels given by the experimental data.

\begin{figure}
\includegraphics[width=130.0mm]{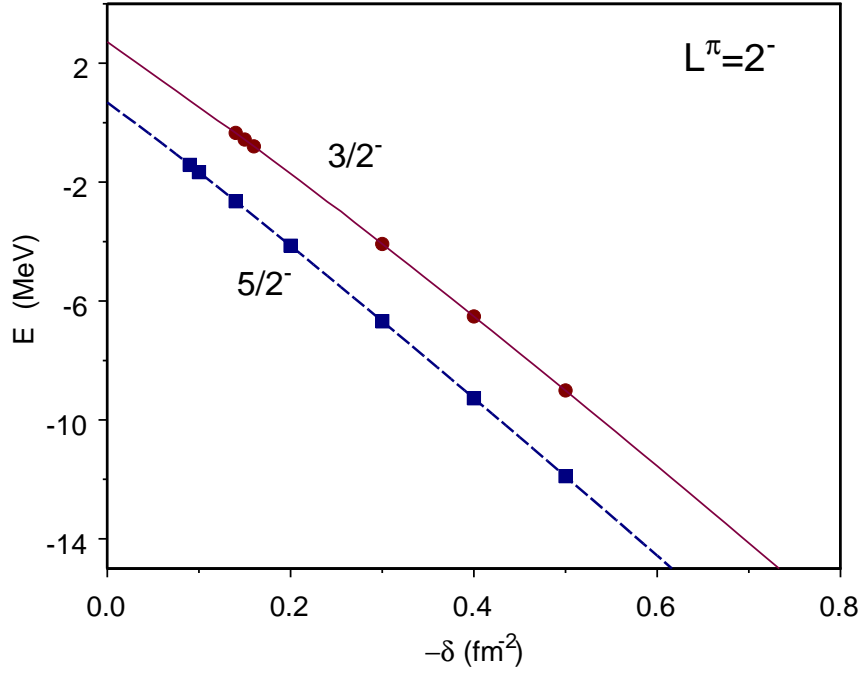}
\caption{\label{Filikhin_fig:3}
The P\'ade approximants  for  the spin-flip doublet ($\frac32^-$, $\frac52^-$)
of  $^9$Be.
Total orbital momentum and parity of each level are $2^-$.  Bound state Energies  calculated for several values of $\delta$ (see Eq.(\ref{3bf})) are depicted by the circles and squares for  the $\frac32^-$ and $\frac52^-$, respectively. Real parts of the Pade approximants are shown by lines. Resonance energy is found at $\delta$=0
(Here we used the value $\hbar^2/2m$=41.44~MeVfm$^2$ for conversion of  MeV to fm$^{-2}$).
Energy is measured from $\alpha+\alpha+n$ threshold.
}
\end{figure}

\begin{figure}
\includegraphics[width=130.0mm]{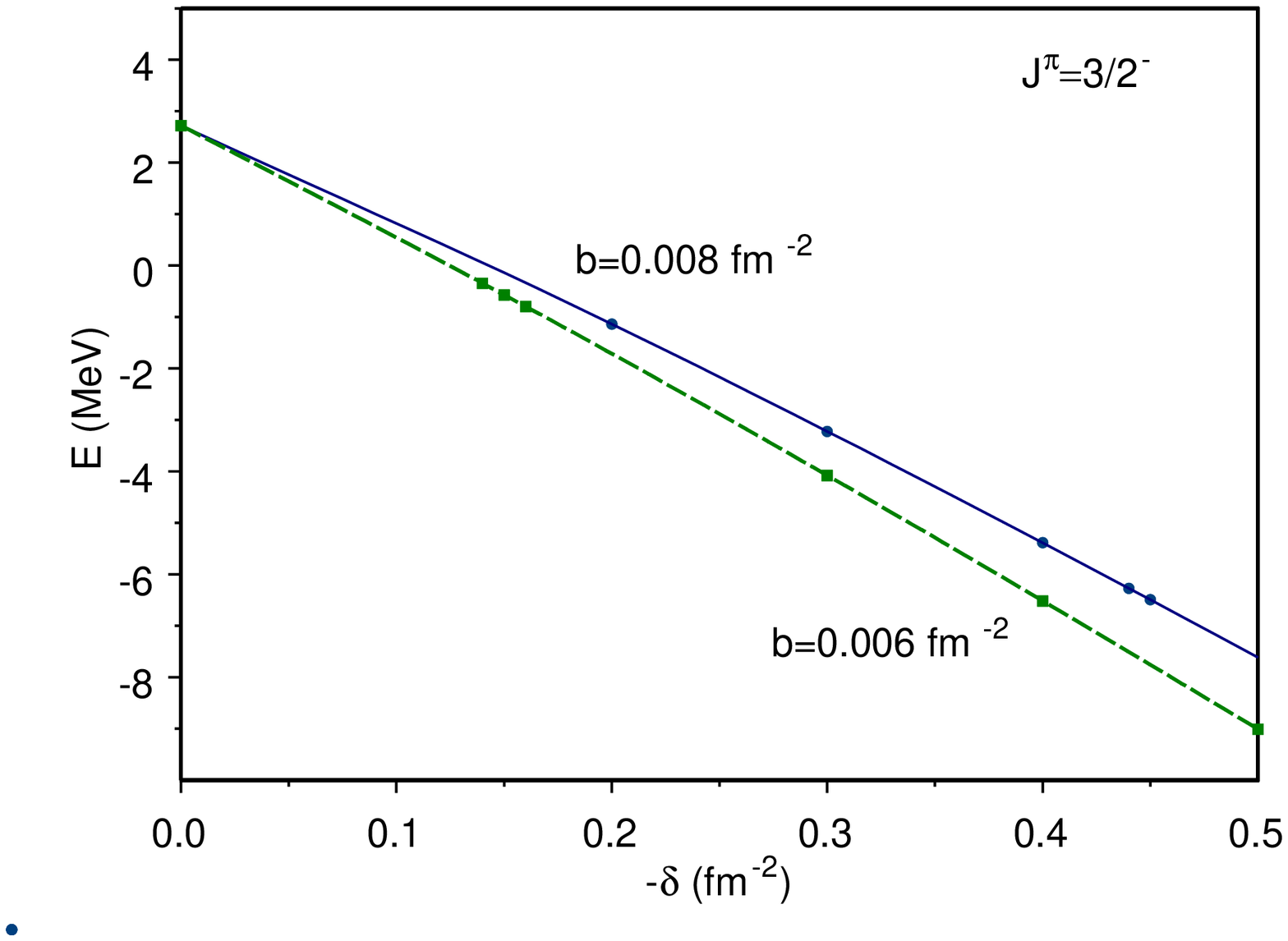}
\caption{\label{Filikhin_fig:4}
The  P\'ade  approximants  for  the the upper member $\frac32^-$ of the $L=2^-$ state
of  $^9$Be. The calculation of resonance energy was performed for two values of the range parameter $b$
of the tree-body potential (\ref{3bf}).  All notations are the same as in Fig. \ref{Filikhin_fig:3}.
}
\end{figure}

Finally, we illustrate our calculations in  Fig. \ref{Filikhin_fig:3}  and \ref{Filikhin_fig:4}. In the first one,
the real parts of the P\'ade approximants  for  the ($\frac32^-$, $\frac52^-$) spin flip doublet
of  $^9$Be are shown. The total orbital momentum and parity of each level are $2^-$. Calculations for several values of the strength parameter $\delta$ (see Eq.(\ref{3bf}))
give a "trajectory" of negative energies (noted by circles in Fig. \ref{Filikhin_fig:3}). Real parts of the constructed P\'ade approximants are shown by lines.
Resonance energies of both levels correspond to the zero value of the argument $\delta$  ($\delta$ =0). From Fig. \ref{Filikhin_fig:3} one can see that
the behavior of the P\'ade approximants  is close to linear dependence. The energies may be determined just as well by eye as by calculation due to the linear behavior
of the P\'ade approximants.

In Fig.  \ref{Filikhin_fig:4} the P\'ade  approximants  for  the $\frac32^-$ upper member of the $L=2^-$ state
is shown for  two values of the range parameter $b$ of the three-body potential (\ref{3bf}). The real parts of the obtained P\'ade approximants are crossed at point $\delta$ =0.
It is a test for our calculations. We have obtained from Fig. \ref{Filikhin_fig:4} that the non-physical three-body potential with different values for the parameter $b$  gives different P\'ade approximants. However the calculated resonance energy does not depend on this parameter. We can conclude that the value obtained for resonance energy  corresponds to the physical value. Non-physical solutions, which are possible using this method, can be separated by this  test.

\section{Summary}
The configuration space Faddeev equations were applied to calculate the energy spectrum of the  $^9$Be
nucleus within the $\alpha+\alpha+n$ cluster model.
We found the set of local phenomenological potentials which accurately reproduce the ground state binding energy and reasonably reproduce the energies of low-lying resonances.
This set includes the Ali-Bodmer potential of the model "e" for $\alpha\alpha$
and a new  $\alpha n$ potential, which was constructed to reproduce the R-matrix fit for the $\alpha n$ scattering data.
In our model the total orbital momentum is fixed for each
energy level. Thus, the $^9$Be energy levels can be classified as members of the spin-flip
doublet corresponding to the total orbital momentum ($L^\pi=0^+, 2^+,4^+, 1^-, 2^-,3^-, 4^-$) of the
system. In the framework of the model, the predictions for the resonance energies of the 4$^+$ and 4$^-$ spin flip-doublets
have been made.

\newpage

\begin{table}
\caption{ Parameters of the $\alpha\alpha$ (ABe \cite{ABo66}) and
$\alpha n$ potentials. The pair orbital momentum is
$l$.  $V_{\rm att}^l$ ($V_{\rm rep}^l$) and $V_{c}$ are given in MeV,
 $\beta_{\rm att}^l$ ($\beta_{\rm rep}^l$) in fm$^{-1}$ and
 $\alpha_{c}$ in
 fm$^{-2}$.
}
\label{tab:1}
\begin{tabular}{clllccc}
\hline\noalign{\smallskip}
  Interaction &    Potential           & $l$ &
$V_{\rm rep}^l$  & $\beta_{\rm rep}^l$ & $V_{\rm att}^l$ & $\beta_{\rm att}^l$ \\
\noalign{\smallskip}\hline\noalign{\smallskip}
$\alpha\alpha$& central            &0 & 1050 & 0.8 & 150 & 0.5 \\
              &                    &2 & 640  & 0.8 & 150  & 0.5 \\
              &                    &4 & --    & --   & 150  &  0.5 \\
\noalign{\smallskip}\hline
$\alpha n      $&central     & 0 \cite{Zhukov1993} &  50.0   &  1/2.3 & -- & -- \\
               &                       & 1 &  40.0   & 1/1.67   & 63.0 & 1/2.3 \\
               &              & 2 \cite{FJR}& --    &  --  & 21.93 & 1/2.03\\
               &spin-orbit  & -- & -- & -- &   38.0   & 1/1.67 \\
\noalign{\smallskip}\hline
\end{tabular}
\end{table}

\newpage

\begin{table}
\caption{Binding energy $E_B$ of the $^{9}$Be ground state ($3/2^-$) (in MeV), calculated
for various orbital momentum configurations.
Energy is measured with respect to the ${\alpha+\alpha+ n}$
threshold. Experimental value is $E^{ex}_B$=-1.5735  MeV \cite{tunl}. }
\label{Filikhin_tab:2}
\begin{tabular}{llc}
\hline
\{($l_{\alpha n}$,$\lambda_{(\alpha n)-\alpha}$)\} & \{($l_{\alpha\alpha}$,$\lambda_{(\alpha\alpha)-n}$)\}& $E_B$ \\
\hline
 (0,1)                               & (0,1) & unbound \\
 (0,1)(1,0)(1,2)(2,1)                &       & unbound\\
 (0,1)(1,0)(1,2)(2,1)(2,3)           &       & -0.032\\
 (0,1)(1,0)(1,2)(2,1)(2,3)(3,2)      &       & -0.042\\
 (0,1)                               & (2,1) & unbound \\
 (0,1)(1,0)(1,2)(2,1)(2,3)(3,2)      &       & unbound \\
                                     & (0,1)(2,1)  & -1.403\\
                                     & (0,1)(2,1)(2,3)  & -1.480\\
                                     & (0,1)(2,1)(2,3)(4,3)  & -1.492\\
                                     & (0,1)(2,1)(2,3)(4,3)(4,5)  & -1.493\\
 \hline
\end{tabular}
\end{table}

\newpage

\begin{table}
\caption{
Energy levels in the $\alpha\alpha n $ system and low-lying $^9$Be spectrum. Results of our calculations
are presented in the third column. The energy (in MeV) is measured from the
${\alpha+\alpha+n}$ threshold. Experimental data for $^9$Be ($T=1/2$) are taken from \cite{tilley}.
}
\label{Filikhin_tab:3}
\begin{tabular}{lcccccl}
\hline
 $L^\pi$ & $J^\pi$ &     &\cite{Arai} (CSM)& \cite{ARJGF}& \cite{P}* &  Exp.  \\
\hline
0$^+$ & $ \frac12^+$ & 0.3(4)  & --  & -- &-- &  0.111$\pm 0.007$ \\ \hline
1$^-$ &$\frac32^-$   & -1.493  & -2.16 & -1.60 & -1.48(4) & -1.5735   \\
      &$\frac12^-$   &  0.9(0)  & 1.06 & --  & 2.24(6) &1.21$\pm 0.12$   \\ \hline
2$^-$ & $\frac52^-$  & 0.7(2)  & 0.39 & --   &0.73(6) &0.8559$\pm 0.0013$  \\
      & $\frac32^-$  & 2.(7)   & 2.88 & 2.65 &-- &4.02$\pm 0.1$ or 3.4$\pm$0.5\cite{Prezado} \\ \hline
2$^+$ & $\frac52^+$  & 1.(8)   & 1.75 & --   &-- &1.476$\pm 0.009$\\
     & $\frac32^+$   & 3.(0)   & 3.21 & --   &-- &3.1312$\pm 0.025$ \\ \hline
3$^-$ & $\frac72^-$  & 4.(6)   & 5.02 &--    &5.03(6) &4.81$\pm 0.06$ \\
     & $\frac52^-$   & 6.(4)   & 6.57 & --   &-- &6.36$\pm 0.08$ \\ \hline
4$^+$ & $\frac92^+$  & 5.(1)   & 5.04 & --    &-- &5.19$\pm 0.06$ \\
     & $\frac72^+$   & 6.(6)   & 6.80 & --   &-- &-- \\ \hline
4$^-$ & $\frac92^-$  & 8.(6)   & 9.73 & --   &8.62(6) &--  \\
     & $\frac72^-$   & 10.(7)  & --   & -- &-- &--  \\
\hline
\end{tabular}\\
\noindent *Quantum Monte Carlo calculations for $A$=9.
\end{table}

\begin{table}
\caption{\label{Filikhin_tab:3}
Results of our calculations for several energy levels in the $\alpha\alpha p $ system and low-lying $^9$B spectrum.
The energy $E$ (in MeV) is measured from the
${\alpha+\alpha+p}$ threshold. The Coulomb energy $\Delta_c$ of proton displacement in $\alpha\alpha n $ system (in MeV) of each level is presented.
Experimental data (denoted as "Exp.") for $^9$B ($T=1/2$) are taken from \cite{tilley}.
}
\begin{tabular}{ccllll}
\hline
 $L^\pi$ & $J^\pi$ &   $E$ & $\Delta_c$ &$E^{Exp.}$ &$\Delta^{Exp.}_c$  \\
\hline
1$^-$ &$\frac32^-$   &  0.1(6)& 0.4 & 0.277             & 0.56   \\
      &$\frac12^-$   &  2.(4) & 0.1 &3.027$\pm 0.3$     & 0.25  \\
\hline
2$^-$ & $\frac52^-$  &  2.(4) & 0.4 & 2.6386$\pm 0.005$ & 0.49\\
\hline
\end{tabular}
\end{table}

\begin{figure}
\includegraphics[width=120.0mm]{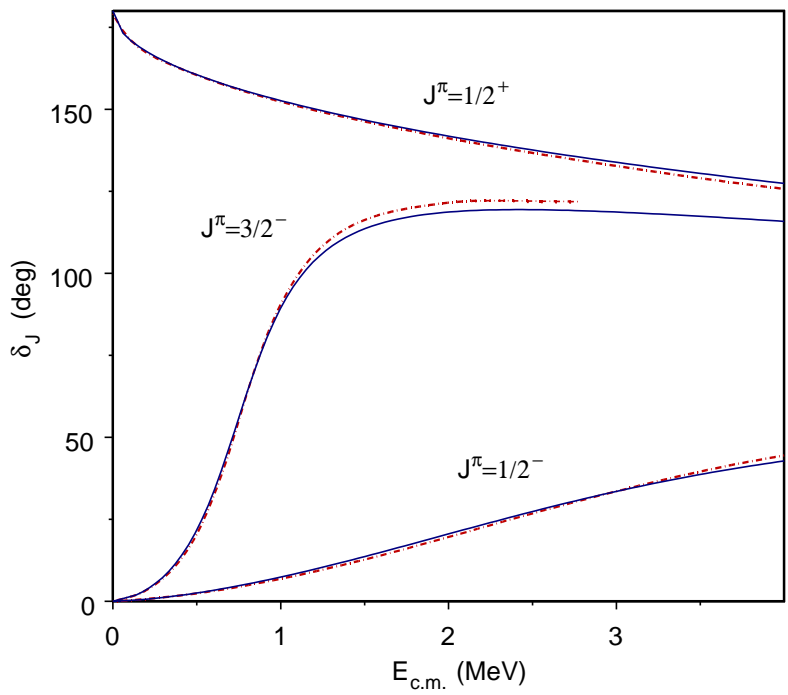}
\caption{\label{Filikhin_fig:2}
Phase shifts for $\alpha-n$ scattering. Solid curves are our results with potential (\ref{an})
and dot-dashed curves are an R-matrix fit to data from \cite{Noll2007}.
}
\end{figure}

\newpage

\begin{thebibliography}{99}
\bibitem{Tang} Y. C. Tang, F. C. Khanna, R. C. Herndon and K. Wildermuth,
Nucl. Phys. 35 (1962) 421.
\bibitem{Arai} K. Arai, P.Descouvemont, D. Baye, W. N. Catford, Phys. Rev. C 68 (2003) 014310.
\bibitem{ARJGF} R. ?lvarez-Rodr?guez, A. S. Jensen, E. Garrido, and D. V. Fedorov, Phys. Rev. C 82 (2010) 034001.
\bibitem{T2006} M. Theeten D. Baye and P. Descouvemont, Phys. Rev. C 74 (2006) 044304.
\bibitem{T2007} M. Theeten, H. Matsumura, M. Orabi, D. Baye, P. Descouvemont, Y. Fujiwara, and Y. Suzuki
Phys. Rev. C 76 (2007) 054003.
\bibitem{Hiyama} E. Hiyama, M. Kamimura, T. Motoba, T. Yamada,
Y. Yamamoto,  Phys. Rev. C 66 (2002) 024007.
\bibitem{SBB} S. Sack, L. C. Biedenharn, G. Bret, Phys. Rev.
93 (1954) 321.
\bibitem{Zhukov1993} M.V. Zhukov, B.V. Danilin, D.V. Fedorov, J.M. Bang, I.J.
Thompson, and J.S. Vaagen, Phys. Rep. 231 (1993) 151.
\bibitem{FJR} D. V. Fedorov, A. S. Jensen, and K. Riisager, Phys. Rev. C 49 (1994) 201; A. Cobis,
D. V. Fedorov and A. S. Jensen, Nucl. Phys. A 631 (1998) 793.
\bibitem{ABo66} S. Ali and A. R. Bodmer,  Nucl. Phys. 80 (1966) 99.
\bibitem{Noll2007} K. M. Nollett, S.C. Pieper, and R. B. Wiringa,
J. Carlson and G. M. Hale, Phys. Rev. Lett.  99 (2007) 022502.
\bibitem{Faddeev}  L. D. Faddeev and S. P. Merkuriev,    {\it Quantum
Scattering Theory for Several Particle Systems} (Kluwer Academic, Dordrecht, 1993) pp. 398.
\bibitem{FGS04} I. Filikhin, A. Gal, V. M. Suslov,
Nucl. Phys. A 743 (2004) 194.
\bibitem{FSV} I. Filikhin, V. M. Suslov and B. Vlahovic, J. Phys. G: {\bf 30}, 513 (2004).
\bibitem{BSSV96} J. Bernabeu, V. M.  Suslov, T. A. Strizh, S. I. Vinitsky,
 Hyperfine Interaction {\bf 101/102}, 391 (1996).
\bibitem{KKH} V. I. Kukulin, V. M. Krasnopolsky and J. Horacek,   {\it Theory of
Resonances} (Kluwer Academic, Dordrecht, 1989).
\bibitem{KK} C. Kurokawa  and K. Kato,  Phys. Rev. C 71, 021301(R) (2005).
\bibitem{FSV05} I. Filikhin, V. M. Suslov and B. Vlahovic, J. Phys. G  31 (2005) 1207.
\bibitem{FSV09} I. Filikhin, V. M. Suslov and B. Vlahovic, Phys. Atom. Nucl.  72  (2009) 619.
\bibitem{Car} R. Lazauskas, J. Carbonell, Phys. Rev. C 71 (2005) 044004.
\bibitem{FJ} D.V. Fedorov, A.S. Jensen,   Phys. Lett.  B  389 (1996) 631.
\bibitem{tunl} Nuclear data evaluation project, http://www.tunl.duke.edu/nucldata/
\bibitem{tilley} D.R. Tilley, J.H. Kelley, J.L. Godwin, D.J. Milliner, J.E. Purcell, C.G. Sheu, H.R. Weller,
Nucl. Phys. A 745 (2004) 155; http://www.tunl.duke.edu/nucldata/
\bibitem{P} S. C. Pieper, K. Varga and R. B. Wiringa1, Quantum Monte Carlo calculations of A=9,10 nuclei, Phys. Rev. C 66 (2002) 044310.
\bibitem{Prezado} Y. Prezado, et al., Phys. Lett. B 618 (2005)43.
\end{thebibliography}
\end{document}